\documentclass[twocolumn,showpacs,preprintnumbers,aps,amsmath,amssymb,prl]{revtex4}

\usepackage{graphicx}
\begin{document}
\title{Observation of Weak-Limit Quasiparticle Scattering via Broadband Microwave Spectroscopy
of a $d$-Wave Superconductor}
\author{P. J. Turner}
\author{R. Harris}
\author{Saeid Kamal}
\author{M. E. Hayden}
\altaffiliation[Permanent address:]{Dept. of Physics, Simon Fraser University,
Burnaby, B.C., Canada, V5A 1S6.}
\author{D. M. Broun}
\author{D. C. Morgan}
\author{A. Hosseini}
\author{P. Dosanjh}
\author{G. Mullins}
\author{J. S. Preston}
\altaffiliation[Permanent address:]{Dept. of Physics and Astronomy, McMaster
University, Hamilton, ON, Canada, L8S 4M1.}
\author{Ruixing Liang}
\author{D. A. Bonn}
\author{W. N. Hardy}
\affiliation{Department of Physics and Astronomy, University of British Columbia,
Vancouver, B.C., Canada  V6T 1Z1}
\date{\today}
\begin{abstract}
There has long been a discrepancy between microwave conductivity measurements
in high temperature superconductors and the conductivity spectrum expected in
the simplest models for impurity scattering in a d-wave superconductor. Here we
present a new type of broadband measurement of microwave surface resistance
that finally shows some of the spectral features expected for a $d_{x^2-y^2}$
pairing state. Cusp-shaped conductivity spectra, consistent with weak impurity
scattering of nodal quasiparticles, were obtained in the 0.6-21~GHz frequency
range in highly ordered crystals of YBa$_2$Cu$_3$O$_{6.50}$ and
YBa$_2$Cu$_3$O$_{6.99}$.
\end{abstract}
\pacs{74.25.Nf, 74.72.Bk}
\maketitle

It is now widely accepted that cuprates exhibiting high temperature
superconductivity involve a pairing state with $d_{x^2-y^2}$ symmetry
\cite{tsueikirtley}. However, there is considerable debate over the extent to
which conventional Bardeen-Cooper-Schrieffer theory can account for the
properties of this superconducting state, a debate that is closely related to
the difficulty in understanding the peculiar metallic state in these materials.
The question is, despite the new physics being encountered in the cuprate phase
diagram, is the $d_{x^2-y^2}$ superconducting state simple? In this letter we
present the first low frequency conductivity spectra that answer this in the
affirmative, showing an evolution at low temperatures towards the cusp-shaped
spectrum expected for weak-limit impurity scattering in a clean $d$-wave
superconductor. This characteristic spectrum has remained elusive because it
only appears at temperatures far below $T_c$ in very clean crystals, thus
necessitating the development of a sensitive bolometric technique for making
broadband microwave measurements.

A simple feature of a clean $d_{x^2-y^2}$ superconductor is that the low energy
density of states is expected to have the form
$N(\varepsilon)\propto\varepsilon/\Delta_0$ where the linear energy dependence
comes from the linear dispersion of the superconducting gap near its nodes in
momentum-space and $\Delta_0$ is the energy scale of the superconducting gap.
At sufficiently low temperatures the microwave conductivity should be governed
by the quasiparticle excitations near the nodes. The microscopic details of the
scattering mechanism and the framework for calculating the conductivity provide
an opening where one might find novel physics entering into the problem. In the
low temperature limit, the conventional framework for calculating the
conductivity in the superconducting state is a self-consistent $t$-matrix
approximation (SCTMA) developed to account for impurity pair-breaking effects
which modify both $N(\varepsilon)$ and the electron self-energy
\cite{HPS,scharnberg}. This work showed that the expression for the microwave
conductivity for any scattering strength takes a simple energy-averaged Drude
form
$\sigma(\omega,T)=ne^2/m^\star\left<[\textrm{i}\omega+\tau^{-1}(\varepsilon)]^{-1}
\right>_{\varepsilon}$, where the energy dependence of the scattering rate
$\tau^{-1}(\varepsilon)$ is determined by the strength of the scattering and
$<\cdots>_{\varepsilon}$ denotes a thermal average weighted by $N(\varepsilon)$
\cite{HPS}. At low $T$, the unitary (strong-scattering) limit result has
$\tau^{-1}(\varepsilon) \approx \Gamma_u / \varepsilon$ while the Born limit
has $\tau^{-1}(\varepsilon) \approx \Gamma_B\varepsilon$ (to within logarithmic
corrections), where the scale factors $\Gamma_u$ and $\Gamma_B$ are determined
by the density of impurities. The puzzling failure of this theory has been that
neither form resembles the energy-independent $\tau^{-1}$ inferred from the
microwave measurements on YBa$_2$Cu$_3$O$_{6.99}$ by Hosseini~\textit{et al.}
\cite{hosseini}.

Two major technical steps have now led to the discovery of a low temperature
regime where some of the expectations of the simple theory seem to hold. First,
the samples used in the experiment reported here are of exceptionally high
crystalline perfection. Most cuprate materials retain a substantial level of
intrinsic disorder due to cation cross-substitution, but the
YBa$_2$Cu$_3$O$_{6+x}$ system has a chemical stability that guarantees an
extremely low level of cation disorder ($<$10$^{-4}$). Large improvements in
the purity of YBa$_2$Cu$_3$O$_{6+x}$ crystals were brought about by the advent
of BaZrO$_3$ crucibles, which do not corrode during crystal growth \cite{BZO}.
With a very high degree of atomic order in the CuO$_2$ planes, the secondary
disorder effect of the off-plane oxygen atoms becomes important. Since
intercalation of oxygen into the CuO chains in this material (which run along
the crystal $\hat{b}$-axis) act as the reservoir for hole doping, probing the
phase diagram typically involves nonstoichiometry and off-plane disorder.
However, a few highly ordered phases are available in this system. Here we
concentrate on fully-doped YBa$_2$Cu$_3$O$_{7}$, with every CuO chain filled,
and ortho-II ordered YBa$_2$Cu$_3$O$_{6.50}$, with every other CuO chain
filled. The perfection of the fully-doped samples is limited by not being able
to anneal to completely full chains, O$_{6.99}$ being our limit. The
crystalline perfection of the YBa$_2$Cu$_3$O$_{6.50}$ crystals comes from the
high purity which in turn allows the development of very long correlation
lengths of the ortho-II order in the 3 crystallographic directions
($\xi_a$=148~\AA, $\xi_b$=430~\AA, and $\xi_c$=58~\AA) \cite{liangOII}. The
measurements shown here were performed on mechanically detwinned single
crystals of 99.995\% purity with dimensions $ (a\times b\times c)\sim(1\times
1\times 0.010)$ mm$^3$.

The second key step to the results presented here is a measurement technique
based on a bolometric method of detection, which provides a natural way of
covering the microwave spectrum in more detail than is possible with a set of
fixed-frequency experiments. The technique has recently proven useful for
measuring resonant absorption in the cuprates \cite{matsuda}. For the present
work we have improved the sensitivity of the technique to the pW level
necessary for resolving the intrinsic power absorption of a small single
crystal, while at the same time employing an in-situ Ag:Au alloy reference
sample of known surface resistance that calibrates the absorption absolutely.
This is particularly important because it allows compensation for the strong
frequency dependence of the microwave field amplitude caused by standing waves
in the microwave circuit.

\begin{figure}
\includegraphics[width=3.2in]{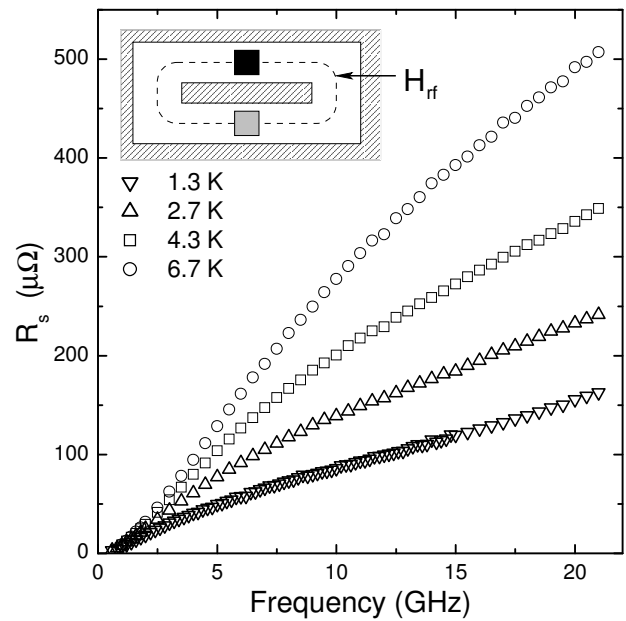}
\caption{\label{fig:Rs} Low temperature surface resistance data for the
$\hat{a}$-axis direction of ortho-II ordered YBa$_2$Cu$_3$O$_{6.50}$ obtained
from synchronous power absorption measurements.  The microwave transmission
line used for the measurements is shown in the inset, with the platelet-shaped
high~$T_c$ (black) and Ag:Au alloy reference (grey) samples positioned in the
regions of homogeneous $\vec{H}_{rf}$. }
\end{figure}

In our system, the microwave surface resistance $R_s(\omega,T)$ is inferred
from the synchronous measurement of the sample temperature rise as the
amplitude of the microwave magnetic field $\vec{H}_{rf}$ is modulated at low
frequency (1~Hz). The high-$T_c$ and reference samples are placed in symmetric
positions in a rectangular shaped coaxial transmission line with a broadened
center conductor designed to ensure spatial magnetic field homogeneity, as
depicted in the inset of Fig.~\ref{fig:Rs}. The coaxial line is terminated with
a short, providing a region where the samples experience predominantly magnetic
fields. Both samples are mounted on thin sapphire plates with a small amount of
silicone grease. Chip heaters mounted on the thermal stages allow precise
calibration of the thermal sensitivity, while the use of a second Ag:Au sample
in place of the high-$T_c$ sample confirms the absolute value and frequency
independence of the calibration. All measurements use the low demagnetization
factor geometry for a thin platelet, with $\vec{H}_{rf}
\parallel \hat{b}$. Thus the screening currents run along the crystal's
$\hat{a}$-axis, completing the loop with a short section of $\hat{c}$-axis
current which we do not attempt to correct for but know to be a small
contribution.

Figure~\ref{fig:Rs} shows the frequency dependence of $R_s$ for a-axis currents
in a crystal of YBa$_2$Cu$_3$O$_{6.50}$. The measurements span 0.6-21~GHz,
limited at low frequency where the small dissipation of the sample approaches
the resolution limit of the experiment. The 1 GHz values for the rms
uncertainty in $R_s$, $\delta R_s$, are about 0.2, 0.4, 0.6, and
1.3~$\mu\Omega$ for T~=~1.3, 2.7, 4.3, and 6.7~K respectfully. For the worst
case of 6.7~K this translates to $\delta \sigma_1$/$\sigma_1$= 0.14 with
uncertainties decreasing correspondingly at higher frequencies and lower
temperatures. For clarity, error bars are omitted from the figures. At high
frequencies, the apparatus is limited by deviations from the TEM field
configuration in the sample region. An overall 5\% uncertainty in $R_s$ is set
mainly by the Ag:Au reference-alloy DC resistivity measurement.

In the limit of local electrodynamics, the relevant limit in our sample
geometry \cite{kosztinleggett}, the surface impedance $Z_s=R_s + \textrm{i}
X_s$ is connected to the complex conductivity $\sigma=\sigma_1-\textrm{i}
\sigma_2$ via $Z_s=\sqrt{\textrm{i} \mu_0 \omega / \sigma}$. Measurements of
both $R_s$ and $X_s$ can be used to determine $\sigma_1$ and $\sigma_2$, but
broadband measurements of only $R_s$ present a problem. A common solution, used
in infrared measurements, is to measure a single quantity such as reflectance
over a wide enough range that Kramers-Kronig and Fresnel equations can be used
to calculate $\sigma_1$ and $\sigma_2$. Another solution is to fit spectra to a
model of $\sigma_1$ and $\sigma_2$ that automatically satisfies the
Kramers-Kronig relations. However, low frequency microwave measurements allow
another option for extracting $\sigma_1(\omega,T)$ from $R_s(\omega,T)$. Deep
in the superconducting state, both $\sigma_2$ and $X_s$ are largely controlled
by the screening associated with the response of the superfluid. In the low
frequency limit $X_s=\mu_0\omega\lambda$ and
$\sigma_2=(\mu_o\omega\lambda^2)^{-1}$ where $\lambda$ is the London
penetration depth. So, at low temperatures a broadband measurement of
$R_s(\omega,T)$ together with a measurement of $\lambda(T)$ at a single
frequency is sufficient to extract the conductivity spectrum
$\sigma_1(\omega,T)$. Using cavity perturbation techniques with a
superconducting loop-gap resonator operating at 1.1~GHz \cite{hardy1993} we
measure $\Delta X_s(T)=\mu_0 \omega \Delta \lambda(T)$, where
$\Delta\lambda(T)=\lambda(T)-\lambda(T=1.2~K)$ is the temperature dependent
increase of the London penetration depth from its low temperature limit. Since
these measurements do not determine $\lambda$ absolutely we take the $T \to 0$
value to be $\lambda_0=1600$~\AA\ for overdoped  YBa$_2$Cu$_3$O$_{6.99}$ as
measured by infrared reflectance \cite{basov} and $\lambda_0=2600$~\AA\ for
underdoped YBa$_2$Cu$_3$O$_{6.50}$, consistent with $\mu$SR measurements
\cite{miller}. A more accurate extraction of $\sigma_1(\omega,T)$ involves a
correction for screening by the normal fluid conductivity. To do this we use a
self-consistent method that includes contributions to $\sigma_2(\omega,T)$ both
from the superfluid and from the screening component due to quasiparticles
\cite{hosseini}. At the low temperatures of this experiment, where there are
few quasiparticles to contribute to the screening, this is a small correction
having a maximum effect of 8\% on $\sigma_1$ for our highest temperatures and
frequencies.

\begin{figure}
\includegraphics[width=3.2in]{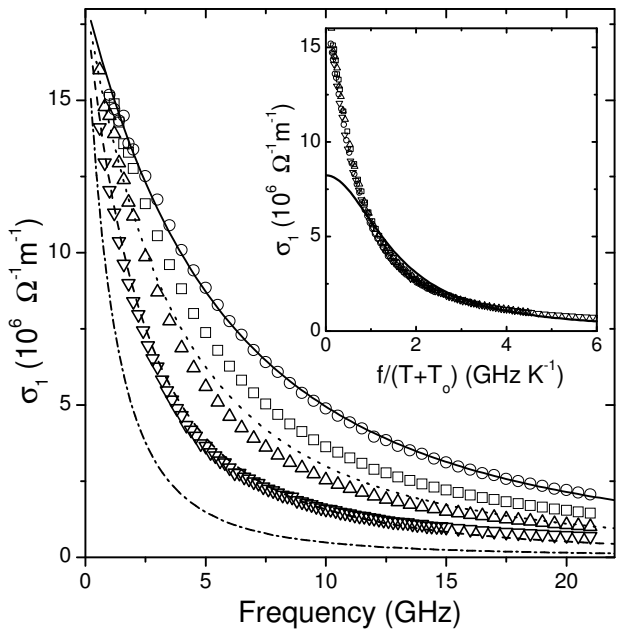}
\caption{\label{fig:born} The low $T$ evolution of the quasiparticle
conductivity spectra of ortho-II YBa$_2$Cu$_3$O$_{6.50}$ (same symbols as Fig.
\ref{fig:Rs}). The solid curve is a fit to the 6.7~K data with the
Born-scattering model, but the dashed curves show that the model fails to
capture the observed T dependence. The inset shows that the data obey an
unusual frequency--temperature scaling
$\sigma_1(\omega,T)=\sigma_1(\omega/[T+T_0])$ with $T_0=2.0$~K. The Drude fit
in the inset illustrates the inadequacy of the Lorentzian lineshape.}
\end{figure}

Figure~\ref{fig:born} depicts the real part of the quasiparticle conductivity
extracted from the broadband $R_s$ measurements. The strong frequency
dependence over intervals as small as 1~GHz (equivalent to a temperature of
0.05~K) shows that the timescale associated with the scattering of low-energy
quasiparticles in these extremely clean samples falls within the bandwidth of
our measurement. The cusp-like shape of $\sigma_1(\omega,T)$, the approximately
$T$-independent low frequency limit and a tail that falls more slowly than
$1/\omega^2$ are all features expected for Born  scattering. The solid curve in
Fig.~\ref{fig:born} shows a convincing fit to the 6.7~K data using the
energy-averaged Drude form with $\tau^{-1}(\varepsilon)=\Gamma_B \varepsilon$,
indicating that the overall shape of the spectrum is well-described by the weak
scattering calculation with fit parameters $\hbar\Gamma_B=0.032$ and
$ne^2\hbar/(m^\star \Delta_0) =5.9\times10^5~\Omega^{-1} \textrm{m}^{-1}$. The
other curves in Fig.~\ref{fig:born} are the Born-limit predictions for the
lower temperatures, using the parameters that fit the 6.7~K data. It is clear
that this model progressively underestimates the spectral weight as $T$ is
reduced, and because of this a global fit for all temperatures produces results
that are less satisfactory. The inset of Fig.~\ref{fig:born} shows that the
$\sigma_1(\omega,T)$ data scale as $\omega/(T+T_0)$ with $T_0$=2.0~K, rather
than scaling as $\omega/T$ as expected in the Born limit. Hence, the weak
scattering limit captures the frequency dependence of the data, but the
requirement of the SCTMA model that the spectral weight vanish as $T \to 0$
leads to disagreement.

\begin{figure}
\includegraphics[width=3.2in]{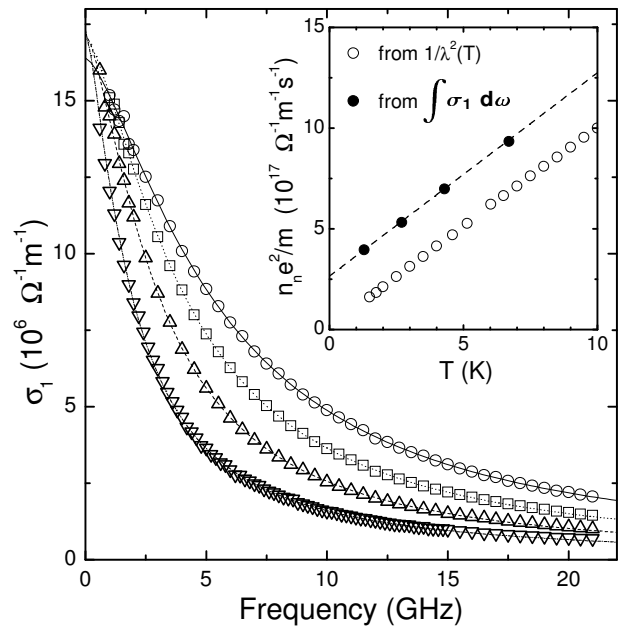}
\caption{\label{fig:phenom} Phenomenological fits with Eq.~\ref{eqn:phenom}
(same symbols as Fig.~\ref{fig:Rs}). The inset compares the normal fluid
density $n_n e^2/m^\star$ obtained by integrating Eqn.~\ref{eqn:phenom} with
the loss of superfluid density inferred from 1.1~GHz $\Delta \lambda(T)$ data.
The dashed line has the slope of the open symbols, and the agreement indicates
that the normal-fluid and superfluid spectral weight obey the conductivity sum
rule, although a residual normal fluid term is implied.}
\end{figure}

In order to fit to the spectra, we adopt a phenomenological form for
$\sigma_1(\omega,T)$ that captures the Born-lineshape features, namely the
cusp-like shape and high frequency tail:
\begin{equation}
\sigma_1(\omega,T)=\sigma_0 / \left[ 1+(\omega / \Gamma)^y \right] .
\label{eqn:phenom}
\end{equation}
We fit directly to $R_s(\omega,T)$, thereby automatically
accounting for the quasiparticle screening. Fig.~\ref{fig:phenom}
shows the fits to individual spectra using this model where the
parameters $\sigma_o$ and \textrm{y} remain relatively constant,
taking average values of $1.67 (\pm 0.05)
~\times~10^7~\Omega^{-1}m^{-1}$ and $1.45 (\pm0.06) $
respectively. The parameter $\Gamma$ varies approximately linearly
in $T$ with fit values $12.1$, $19.2$, $26.3$, and $35.0 \times
10^{9} sec^{-1} $. Although these fits also suggest the unusual
frequency--temperature scaling discussed previously, enforcing it
in the model reduces the agreement in the spectral weight
comparison.  The fits can be integrated in order to obtain {\it
absolutely} the temperature dependent spectral weight associated
with the normal fluid. In the inset of Fig.~\ref{fig:phenom} this
normal fluid spectral weight is compared to the spectral weight
lost from the superfluid as determined independently by the
measurements of $\Delta\lambda(T)$. The slopes agree to within
2\%, a good indication that the model is capturing the oscillator
strength at higher frequencies.  It is important to note that at
low $T$ this comparison is independent of the choice of
$\lambda_0$ to first order since $\sigma_1 \approx 2 R_s/(\mu_0^2
\omega^2 \lambda_0^3)$ and $\Delta(1/\lambda^2) \approx - \Delta
\lambda/\lambda_0^3$. The offset apparent in the normal fluid
density corresponds to a $T=0$ residual normal fluid.

\begin{figure}
\includegraphics[width=3.2in]{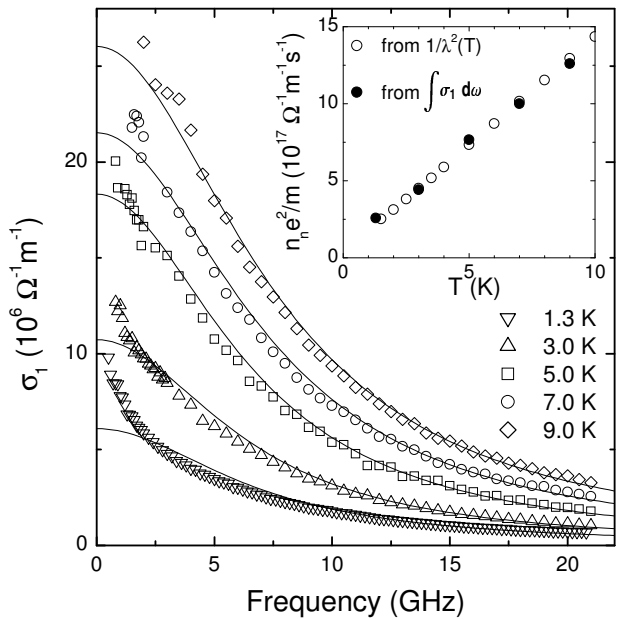}
\caption{\label{fig:orthoI} The conductivity spectrum of a fully-doped sample
of YBa$_2$Cu$_3$O$_{6.99}$ in the $\hat{a}$-direction. The Drude fits to the
spectra highlight the evolution from a cusp-like shape to a more Lorentzian
lineshape with increasing temperature.}
\end{figure}

The discovery of a cusp-shaped spectrum for the low temperature conductivity of
YBa$_2$Cu$_3$O$_{6.50}$ seems at odds with the Drude fits ($y=2$ in
Eq.~\ref{eqn:phenom}) that were found to reasonably describe the spectra
inferred from fixed-frequency microwave measurements on fully-doped
YBa$_2$Cu$_3$O$_{6.99}$ \cite{hosseini}. To resolve this conflict, broadband
conductivity spectra were obtained for the $\hat{a}$-axis of a crystal of the
fully doped material, shown in Fig.~\ref{fig:orthoI}. At the lowest
temperatures there is indeed a cusp-like spectrum, similar to the one seen in
the underdoped ortho-II sample. The reason this was not seen in earlier
measurements is that the spectrum gives way to a more Lorentzian lineshape
above 4~K, as indicated by the progressively better fit to a Drude model with
increasing temperature; the 5 spectra do not scale in the manner seen for
YBa$_2$Cu$_3$O$_{6.50}$. Hosseini~\textit{et al.} concluded that the scattering
rate was nearly temperature independent below 20~K at a value of
$\tau^{-1}=5.6(\pm0.6) \times 10^{10}~sec^{-1}$, and here we find
$\tau^{-1}=4.4(\pm0.3) \times 10^{10}~sec^{-1}$ with the decrease likely due to
continued improvements in the sample purity. The integration of the Drude model
captures the oscillator strength rather well, except at the lowest temperatures
where the fits are too poor for us to comment upon an extrapolated residual
value. It is not yet clear why the non-Lorentzian lineshapes are restricted to
such low temperatures in the fully doped sample.

To conclude, we have provided the first highly detailed measurement of the
quasiparticle conductivity spectrum in the disorder-dominated regime of an
extremely clean $d$-wave superconductor. This regime has been accessed by using
well-ordered crystals with very high purity, measured with a broadband
microwave technique whose frequency range matches the very small quasiparticle
scattering rate in the samples. A number of puzzles remain, particularly the
residual normal fluid inferred from extrapolations to $T=0$, a phenomenon that
is seen to a much greater degree in other cuprates \cite{broun,orenstein}. The
cause of the evolution from a cusp-like spectrum to a more Lorentzian lineshape
in YBa$_2$Cu$_3$O$_{6.99}$ is also unclear and may require a more sophisticated
scattering model that employs intermediate scattering strengths, neither Born
nor unitary. Nevertheless, several features of the quasiparticle scattering
dynamics are characteristic of Born-limit scattering, namely a cusp-shaped
conductivity spectrum seen at low temperatures. Additional physics such as
order parameter suppression at impurity sites \cite{hettlerhirschfeld} might be
needed to resolve the remaining puzzles, but the data presented here provide a
simple starting point that is quite close to the expectation for nodal
quasiparticles scattered weakly by impurities.

\acknowledgments We gratefully acknowledge useful discussions with A.J.
Berlinsky, C. Kallin, and P.J. Hirschfeld, as well as financial support from
the Natural Science and Engineering Research Council of Canada and the Canadian
Institute for Advanced Research.

\end{document}